\documentclass[12pt]{emulateapj}

\begin{document}

\title{A new way to infer variations of the seismic solar radius}


\author{I. Gonz{\'a}lez Hern{\'a}ndez}
\affil{ National Solar Observatory\footnote
	{Operated by the Association of Universities for Research in 
	Astronomy, Inc. under cooperative agreement with the National 
	Science Foundation.}, \\ 950 N. Cherry Ave., Tucson, AZ, USA}
\author{P. Scherrer}
\affil{Stanford University, Stanford, Ca, USA}
\author{F. Hill}
\affil{ National Solar Observatory, 950 N. Cherry Ave., Tucson, AZ, USA}


\email{irenegh@nso.edu}

\begin{abstract}

We show that the mean phase of waves propagating all the way from the far side of the Sun to the front side, as measured by seismic holography, varies with time. The change is highly anticorrelated with solar cycle activity and is consistent with other recent results on the variation of the seismic radius of the Sun. The phase change that we observe corresponds to a few kilometers difference in the seismic solar radius from solar maximum to solar minimum in agreement with inferrences from global helioseismology studies.

\end{abstract}


\section{Introduction}
Accurate knowledge of the solar radius is a key factor in structure inversions from solar-oscillation frequencies. Although several values of the photometric radius have been reported by different authors, the accepted radius for most inversions has been the revised one by \citet{brown1998}.
Measurements of temporal variation of the photometric solar radius are largely controversial. Different techniques have been used to determine the  variations and the results differ, ranging from no variation, to rapid variations, to solar cycle variations. \citet{kuhn2004} ascribed the apparent disparities in the measured optical solar diameter from ground-based instruments to the changing terrestrial atmosphere. They suggest that a solar cycle variation of the Earth's atmosphere is responsible for the correlation/anticorrelation between the observable radius and the solar activity cycle.

\citet{schou1997} introduced the concept of the seismic radius of the Sun. They obtained the solar radius  from $f$-mode frequencies and found a discrepancy with the photospheric standard value of $-$300\,km. \citet{takata2001} confirmed the value of the seismic radius from $p$-mode analysis. Recently, \citet{harberreiter2008} explained the discrepancies between the photospheric optical radius and the seismic radius as arising from the difference between the height at disk center and the inflection point of the intensity profile at the limb.

Since the seismic radius is inferred from helioseismology inversions, it does not necessarily agree with the physical radius. Changes in the stratification of the Sun, such as variations of the sound speed or in the subsurface superadiabatic layers, influence its determination. Hence, variations of the seismic radius are better interpreted as a measurement of changes in the cavity in which the waves propagate, rather than a physical change of the solar radius. Moreover, localized perturbations due to magnetic field concentration can also play a role in the determination of the seismic radius.

Recent results from \cite{lefebvre2005} using $f$-modes and \cite{kholikov2008} using very low-$l$ $p$-modes, found a change of the seismic solar radius of a few kilometers that is anticorrelated with the solar magnetic cycle.
Variations of the solar seismic radius could have large implications, from explaining the solar cycle variations of the observed oscillation frequencies to affecting the seismically-inferred surface maps of far-side activity.

Using seismic holography, a local-helioseismology technique that is here applied to waves that travel from the far side to the front side, we find a variation of the seismic radius with the solar cycle that qualitatively agrees with the global results. The use of local-helioseismology techniques for this type of study allows us to discriminate between the variation being a global phenomenon or a localized one associated with surface magnetic activity.

\section{Data Analysis}
Seismic imaging was proposed by \citet{lindsey1990} as a technique to map active regions on the far hemisphere of the Sun. Since then, the method has also been used successfully to explore the subphotosphere of active regions \citep{chou2000,braun2000}, investigate seismic sources from flares \citep{donea2006} and characterize flows in the solar subphotosphere \citep{braun2004}. Phase-correlation seismic holography applied to high quality data from the Michelson Doppler Imager (MDI) aboard the {\it Solar Heliospheric Observatory (SOHO)} rendered the first seismic image of an active region on the far hemisphere of the Sun \citep{lindsey2000a}.

The phase-sensitive holography technique is based on the fact that there is a phase shift (travel delay) between waves entering and exiting an active region \citep{braun1992}. In the analysis, the waves going out from a particular point in the Sun (focus) are characterized by the acoustic egression, $H_{+}$, and the waves going in are characterized by the acoustic ingression, $H_{-}$.

\begin{equation}
H_{\pm}({\bf r}, ~z, ~t)
 ~=~ \int dt' \int_{P_\pm} d^2{\bf r}'
 ~G_{\pm}({\bf r}, ~{\bf r}', ~z, ~t, ~t')
 ~\psi({\bf r}', ~t'),
\label{eq:H}
\end{equation}

where ${\bf r}$ is the horizontal location of the focus, $z$ is its depth, $G_{\pm}$ are Green's functions that represent the disturbance at $({\bf r}, ~z, ~t)$ resulting from a unit acoustic impulse originating at the focus.
For a detailed explanation of the method see \citet{lindsey2000b}. In the particular case of maps of the far side, ${\bf r}$ is located at the surface of the non-visible hemisphere and the signal is measured at the Earth-facing hemisphere in an annular pupil centered with respect to the far-side focus.

The far-side maps calculated from seismic holography represent the phase of the complex correlation between waves going in and out of a particular point on the non-visible hemisphere. In the quiet Sun, the propagation of the waves is well represented by the model Green's functions. However, when the focus is located in an area of high magnetic activity, the measured signal deviates from the Green's functions, effectively introducing a phase shift between the waves going in and out. In the Fourier domain, the correlation becames the product of the Fourier transforms, $\hat H_+ ({\bf r}, ~0, ~\omega)$ and $\hat H_-^* ({\bf r}, ~0, ~\omega)$
\begin{equation}
C({\bf r}) ~\equiv ~\int _{\omega _1}^{\omega _2} d\omega
 ~\hat H_+({\bf r}, ~0, ~\omega)
 \hat H_-^*({\bf r}, ~0, ~\omega),
\label{eq:C}
\end{equation}
and the phase ($\phi ({\bf r}) ~\equiv ~\arg{C({\bf r})}$) is related to the perturbed travel time by
\begin{equation}
\Delta\phi ~=~ - \omega \Delta\tau.
\label{eq:tau}
\end{equation}

\begin{figure}
\includegraphics[scale=0.8]{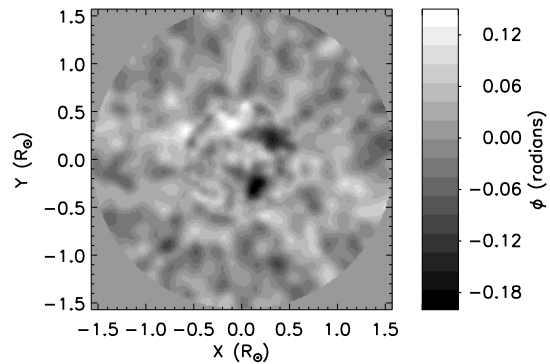}
\caption{\label{fig:postel}Postel projection of the far-side map of $\phi$ for 2003 November 12 showing the strong seismic signatures close to the center (antipode of the Earth-facing hemisphere) of active regions NOAA10488(10507) and NOAA10486(10508).}
\end{figure}

 The sensitivity of the method is not uniform across the disk. This is due to the changing geometry of the front side pupils as the focus move across the far-side hemisphere.
Long-term averages of far-side phase maps are removed from individual maps to correct for this trend. Each far-side map is computed from a one-day series of 1440 Global Oscillation Network (GONG) or Michelson Doppler Imager (MDI) Dopplergrams taken at a cadence of 1 minute. The Dopplergrams are taken in the photospheric line Ni$\,$I~$\lambda$6768$\,$\AA. Each Dopplergram is Postel projected, tracking solar rotation, into a 200$\times$200-pixel map.
Waves following a ray path that bounces once at the solar surface before arriving to the front side (2-skip) are used to map the central part of the far side. A combination of a 1-skip and 3-skip ray paths are used to extend the map over the full far-side hemisphere. Only waves with temporal frequencies between 2.5 and 4.5\,mHz are used for the analysis.
The maps are stacked into a 1440-frame data cube to which the analysis described above is applied. The resulting far-side image is itself a Postel projection. To reduce the errors in the calibration, only maps calculated from Doppler time series with a clear-weather duty cycle greater than 85$\%$ have been considered.

\begin{figure}
\includegraphics[scale=0.7]{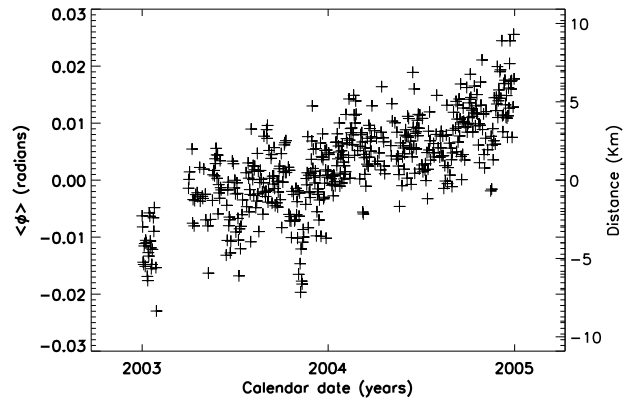}
\caption{\label{fig:phase} Mean phase variation calculated from GONG far-side maps corresponding to the two-years interval 2003-2004. The right hand axis indicates the corresponding extra distance that a wave with a temporal frequency of 3.5\,mHz will travel for the phase to be shifted accordingly to the left hand axis.}
\end{figure}

Figure~\ref{fig:postel} shows an example of a far-side map for 2003 November 12 after correction. Dark areas in the map signify the relatively negative phase signatures, $\phi$, manifested by active regions. In this particular map, active regions NOAA\,10488 and NOAA\,10486 are seen on the far side approximately 7-8 days after they rotated across the West limb of the Sun, disappearing from direct view. They appeared on the near side in the succeeding Carrington rotation. The Postel-projected far-side maps are then reprojected onto a longitude-sin(latitude) grid.

\begin{figure*}
\includegraphics[scale=0.55]{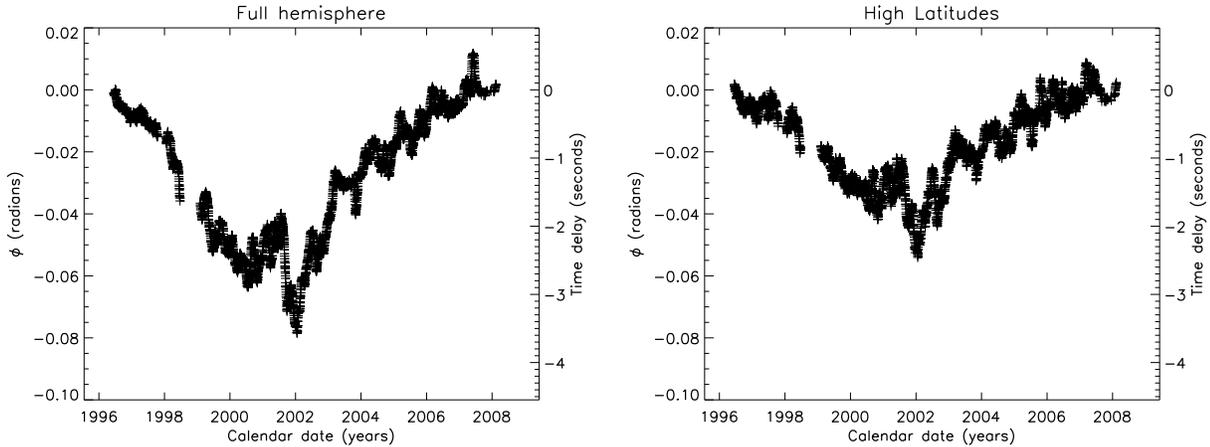}
\caption{\label{fig:radius}Mean phase versus time. Left panel: mean phase calculated over the entire far-side hemisphere. Right panel: mean phase associated to high latitudes only. The right hand axis show the corresponding time delay associated with the phase shift. The change in the mean phase is highly anticorrelated with cycle 23 of solar activity.}
\end{figure*}

Traditionally, the temporal average used to flatten the far-side maps is calculated for a period of low activity. A thorough investigation of the mean phase of the far-side maps over long time periods has produced the unexpected result of a non-constant mean phase. Basically, we find a variation of the quiet Sun from the model Green's functions.
Figure~\ref{fig:phase} show the mean phase of individual farside maps calculated using Global Oscillations Network Group (GONG) data for January 2003 to December 2004 after the average for the whole period has been subtracted. There is a relative variation of the phase of $\approx$\,0.2 radians over the considered time span. A change in the mean phase can be explained by a variation of the cavity in which the waves propagate.

The right-hand axis of Figure~\ref{fig:phase} shows the estimated distance that an acoustic wave with $\nu$\,=\,3.5\,mHz (the central frequency of the wave packets considered) would need to travel at the photospheric sound speed ($\approx$ 8\,Km/s) to account for a particular phase difference. However, as described in the introduction, the change in the seismic radius is not necessarily related to a physical change of the solar radius.
It is important to note that the wave packets considered for calculating far-side maps bounce either none, one or two times before arriving on the front side and cross the subsurface layers several times. Hence, the expansion/contraction of the subsurface layers, or any other effect that produce the change in the seismic radius, could be amplified before the phase displacement is measured.

\section{Solar cycle variation of the mean phase}

In order to search for long term changes of the mean phase, we examine far side maps calculated from MDI Dopplergrams twice daily from May 1996 to July 2008. An average of the complex correlation of individual maps calculated for all of 2007, to minimize contribution from activity, is used to correct for sensitivity change across the disk. Due to this correction, the deviation of the mean phase from zero is relative to the averaged period used for the correction, rather than from the period of time for which the Green's functions for the quiet Sun are optimal.

Once the individual maps have been corrected, the mean phase is calculated for each map. Figure~\ref{fig:radius} presents a 30-day average of the mean phase after correction for the disk trend. The right axis show the corresponding travel time delay as per equation \ref{eq:tau}. The left panel shows the mean phase calculated over the full far-side hemisphere. The right panel shows the mean phase calculated for wave packets originating at high latitudes ($> 60^\circ$).

It is important to notice that the presence of active regions at the far side surface will, by itself, introduce a variation of the mean phase that will be correlated with the solar cycle.
This contribution of the surface activity over the whole Sun is probably responsible for the differences between the mean phase calculated using the full far-side hemisphere (Figure~\ref{fig:radius}, left) and using exclusively the high latitudes (Figure~\ref{fig:radius}, right), where no strong magnetic activity is expected. By choosing the focus of the analysis to be at high latitudes, we are in effect eliminating the {\it in focus} contribution of active regions. However, since the wave packets considered for far-side mapping bounce from one to three times at the surface on their way to the front side, high latitude maps are still affected by {\it out of focus} contribution of active regions. We expect this contribution to be more diffuse. To accurately account for quiet areas only, a through analysis of helioseismic holography maps of higher resolution, where both the focus and the pupils are located on the front side, needs to be carried out.

\begin{figure*}
\includegraphics[scale=0.70]{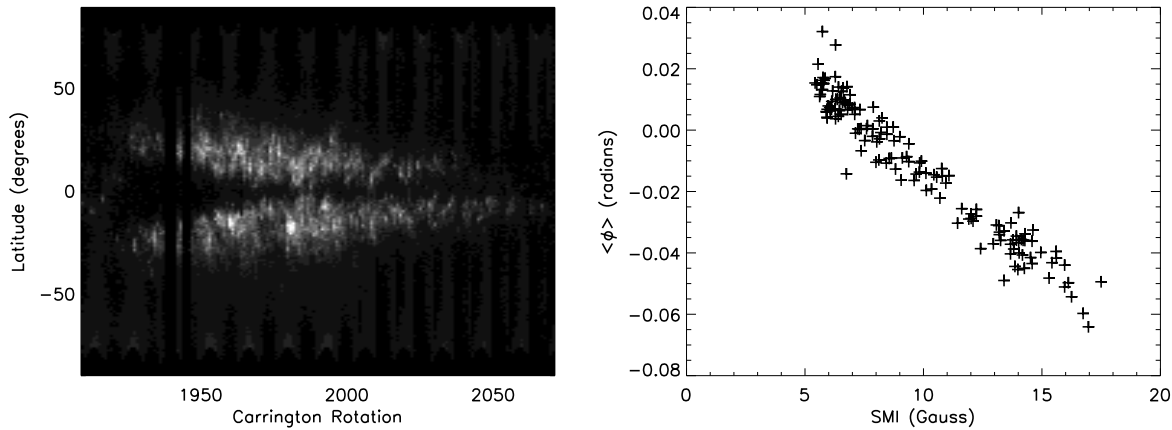}
\caption{\label{fig:butter} Left panel: Butterfly-like diagram showing the position and averaged strengh of the surface activity over time from Carrington rotation 1909 (May 1996) to 2071 (July 2008). Right panel: Scatter plot between the mean phase and the Synoptic Magnetic Index. The mean phase have been averaged over Carrington rotations for consistency with the SMI.}
\end{figure*}

In order to qualitatively check the correlation of the mean-phase variation with the solar cycle, we define a Synoptic Magnetic Index (SMI) as the averaged line-of-sight magnetic-field strengh for individual MDI synoptic magnetograms. The left panel of Figure~\ref{fig:butter} shows the longitudinally averaged line-of-sight magnetic-field strength for the Carrington rotations during the analyzed period (1909-2071). This representation shows the position and strength of the surface activity. We further average the data in latitude for each Carrington rotation to obtain the described SMI. Carrington rotations 1938 to 1941 and 1944 to 1946 were eliminated of the study because of incomplete synoptic magnetograms.

The right panel of Figure~\ref{fig:butter} shows a scatter plot between the SMI and the mean phase obtained for the entire far-side hemisphere that have been averaged over individual Carrington rotations. The strong correlation is clearly seen in the figure.

\section{Discussion and Conclusions}
We find a clear anticorrelation between the seismic radius of the Sun and the solar activity cycle using a local helioseismology technique based in the analysis of wave packets. The results agrees qualitatively with previous work by \citet{lefebvre2005} and \citet{kholikov2008} obtained by analysing global modes.

The term seismic or acoustic radius refers to the wider concept of a cavity within which the waves propagate. In addition to a change of the physical radius, a variation of sound speed, temperature or changes in the superadiabatic superficial layers (surface term), either global or localized to particular depths or latitudes, would influence the seismic radius. Even assuming that the mean phase variation that we find is related to a change of the physical radius, the range of change is well inside the error bars found by \citet{kuhn2004}, confirming that the seismic inferences of the radius are more precise than that of photospheric observations \citep{schou1997}.

Because the inferred variations of the seismic solar radius from helioseismic holography are associated with the propagation of wave packets, rather than individual modes, and the packets bounce from none to two times before reaching the front side where the signal is measured, we need to be careful  interpreting the results in terms of absolute quantities. Wave packets that propagate directly from the far-side focus to the front-side pupil will cross the outer layers two times, while those packets that bounce two times will cross the outer layers six times. This means our results in terms of seismic radius variation need to be divided, on average, by a factor of eight, since we are comparing waves going into and out of the focus, resulting in inferences not far from those of global helioseismology studies \citep{lefebvre2005,kholikov2008}.

The advantage of using local-seismology methods is the ability to infer local properties of the Sun. The present work opens the road to investigate whether the variation of the seismic radius is a global phenomenon or is due to variations associated with the surface magnetic activity. The marked difference between the mean phase variation from waves coming from the poles with those averaged over the full far-side hemisphere (see Figure~\ref{fig:radius}) seems to indicate that waves travelling through areas of high activity contribute more to the mean phase displacement, as expected. Global modes propagate throughout the Sun and are affected by local perturbations. Hence the need of using local helioseismology to separate the relative contributions.
More work is needed to disentangle the contribution of the surface magnetic areas from the apparent global change. Our preliminary attempt to isolate wave packets that have been less affected by surface activity seems to indicate an attenuation of the solar cycle variation in the quiet areas. Higher-resolution local helioseismology has the ability to separate quiet and active areas, which will help to further understand the nature of the variation.
 
As a side effect of the seismic radius variation, the far side maps calculated from seismic holography need to take this temporal change into account and apply the appropriate correctio  throughout the solar cycle in order to stabilize the seismic signal from active regions.

\section*{Acknowledgments}
The authors thank C. Lindsey and D. Braun for their large contribution to MDI and GONG farside pipelines and S. Kholikov, C. Rabello-Soares and J. Fontenla for useful discussions.
MDI Dopplergrams and magnetograms have been using for this investigation. {\it SOHO} is a project of international cooperation between ESA and NASA.
This work utilized data obtained by the Global Oscillation Network Group (GONG) program, managed by the National Solar Observatory, which is operated by AURA, Inc. under a cooperative agreement with the National Science Foundation. The data were acquired by instruments operated by the Big Bear Solar Observatory, High Altitude Observatory, Learmonth Solar Observatory, Udaipur Solar Observatory, Instituto de Astrof{\'{\i}}sica de Canarias, and Cerro Tololo Interamerican Observatory.
This work has been supported by the NASA Living with a Star - Targeted Research and Technology program - and the Stellar Astrophysics branch of the National Science Fundation.


\begin{thebibliography}{}


\bibitem[Braun et al. (1992)]{braun1992}
Braun, D. C., Duvall, T. L., Jr., Labonte, B. J., Jefferies, S. M., Harvey, J. W. \& Pomerantz, M. A. 1992, ApJ, 391, 113

\bibitem[Braun \& Lindsey (2000)]{braun2000}
Braun, D. C., Lindsey, C., 2000, Solar Phys. 192, 307

\bibitem[Braun, Birch \& Lindsey (2004)]{braun2004} Braun, D. C., Birch, A. C., Lindsey, C. 2004, in SOHO 14 / GONG 2004 Workshop. Helio- and Asteroseismology: Towards a Golden Future, ed. Danesy, D., ESA SP-559, 337.

\bibitem[Brown \& Christensen-Dalsgaard (1998)]{brown1998}
Brown, T.M. \& Christensen-Dalsgaard, J. 1998, ApJ, 500, L195

\bibitem[Chou (2000)]{chou2000}
Chou, D.-Y. 2000, Solar Phys. 192, 241

\bibitem[Donea et al. (2006)]{donea2006}
Donea, A.-C., Besliu-Ionescu, D., Cally, P. S., Lindsey, C. \& Zharkova, V. 2006, Solar Phys. ,239, 113

\bibitem[Fazel et al. (2008)]{fazel2008}
Fazel, Z., Rozelot, J.P., Lefebvre, S., Ajabshirizadeh, A. \& Pireaux, S. 2007, New Astronomy, 13-2,65

\bibitem[Gonz{\'a}lez Hern{\'a}ndez et al. (2007)]{gonzalezhernandez2007}
Gonz{\'a}lez Hern{\'a}ndez, I., Hill, F. \& Lindsey, C. 2007, ApJ, 669, 1382

\bibitem[Haberreiter et al. (2008)]{harberreiter2008}
Haberreiter, M., Schmutz, W. \& Kosovichev, A. G. 2008, ApJ, 675, L53

\bibitem[Kholikov \& Hill (2008)]{kholikov2008}
Kholikov, S.and Hill, F. 2008, Solar Phys., 251, 157

\bibitem[Kuhn et al. (2004)]{kuhn2004}
Kuhn, J. R., Bush, R. I., Emilio, M. \& Scherrer, P. H. 2004, ApJ, 613, 1241

\bibitem[Lefebvre \& Kosovichev (2005)]{lefebvre2005}
Lefebvre, S. \& Kosovichev, A. G. 2005, ApJ, 633, L149

\bibitem[Lindsey \& Braun (1990)]{lindsey1990}
Lindsey, C. \& Braun, D. C. 1990, Solar Phys. 126, 101

\bibitem[Lindsey \& Braun (1997)]{lindsey1997}
Lindsey, C. \& Braun, D. C. 1997, ApJ, 485, 895

\bibitem[Lindsey \& Braun (2000a)]{lindsey2000a}
Lindsey, C. \& Braun, D. C. 2000a, Science, 287, 1799

\bibitem[Lindsey \& Braun (2000b)]{lindsey2000b}
Lindsey, C. \& Braun, D. C. 2000b, Solar Phys., 192, 261

\bibitem[Schou et al. (1997)]{schou1997}
Schou, J., Kosovichev, A.G., Goode, P.R. \& Dziembowski, W.A. 1997, ApJ, 489, L197

\bibitem[Takata \& Gough (2001)]{takata2001}
Takata, M. \& Gough, D. O. 2001 in SOHO 10/GONG 2000 Workshop: Helio- and asteroseismology at the dawn of the millennium, ed. Wilson, A., ESA SP-464, 543
 




\end{thebibliography}
\end{document}